\documentclass[prl,twocolumn,floatfix,superscriptaddress,showpacs]{revtex4}

\usepackage{times}
\usepackage{amssymb}
\usepackage[dvips]{graphicx}
\usepackage{epsfig}
\usepackage{bm} 
\usepackage[bf,small]{caption}
\usepackage[bf,small,hang,center,nooneline]{subfigure}

\begin{document}

\title{Analysis of the Scanning Tunneling Microscopy Images of the Charge Density Wave Phase in Quasi-one-dimensional $\mbox{Rb}_{0.3}\mbox{MoO}_{3}$}

\author{E.~Machado-Charry}
\affiliation{Institut de Ci\`{e}ncia de Materials 
de Barcelona (ICMAB--CSIC), 
Campus de la UAB, 08193 Bellaterra,
Spain}
\author{P.~Ordej\'{o}n}
\affiliation{Institut de Ci\`{e}ncia de Materials
de Barcelona (ICMAB--CSIC),
Campus de la UAB, 08193 Bellaterra,
Spain} 
\author{E.~Canadell}
\thanks{To whom correspondence should be addressed; email: canadell@icmab.es}
\affiliation{Institut de Ci\`{e}ncia de Materials 
de Barcelona (ICMAB--CSIC), 
Campus de la UAB, 08193 Bellaterra,
Spain}
\author{C. Brun}
\affiliation{Laboratoire de Photonique et de Nanostructures, CNRS, route de Nozay, 91460 Marcoussis, France
}  
\author{Z. Z. Wang}
\affiliation{Laboratoire de Photonique et de Nanostructures, CNRS, route de Nozay, 91460 Marcoussis, France
}  

\date{\today}

\begin{abstract}
The experimental STM images for the CDW phase of the blue bronze  $\mbox{Rb}_{0.3}\mbox{MoO}_{3}$ have been successfully explained on the basis of first-principles DFT calculations. Although the density of states near the Fermi level strongly concentrates in two of the three types of Mo atoms ($\mbox{Mo}_{\rm II}$ and $\mbox{Mo}_{\rm III}$), the STM measurement mostly probes the contribution of the uppermost O atoms of the surface, associated with the $\mbox{Mo}_{\rm I}\mbox{O}_{6}$ octahedra. In addition, it is found that the surface concentration of Rb atoms plays a key role in determining the surface nesting vector and hence the periodicity of the CDW modulation. Significant experimental inhomogeneities of the $\bm {b^*}$ surface component of the wavevector of the modulation, probed by STM, are reported. The calculated changes in the surface nesting vector are consistent with the observed experimental inhomogeneities. 
\end{abstract}


\maketitle

\begin{figure}
\centering
\subfiguretopcapfalse
\subfigure[]{\hbox to 0.5in{\hfil\null}}%
\setcounter{subfigure}{0}
\subfiguretopcaptrue
\subfigure
{
    \label{topview}
\includegraphics*[height=4cm]{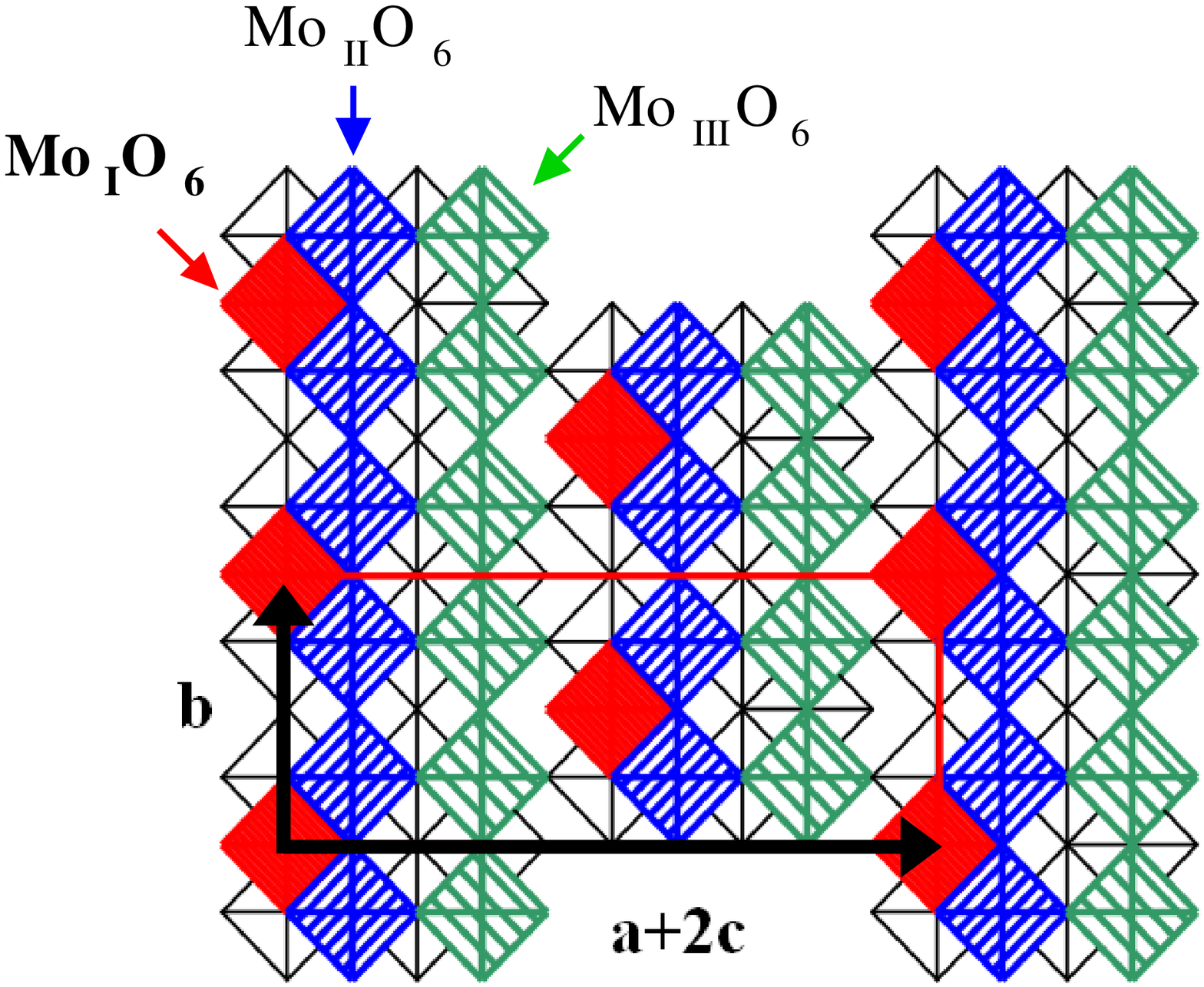}
}
\subfiguretopcapfalse
\subfigure[]{\hbox to 0.7in{\hfil\null}}%
\subfiguretopcaptrue
\setcounter{subfigure}{1}
\subfigure
{
    \label{sideview}
    \includegraphics*[height=4cm]{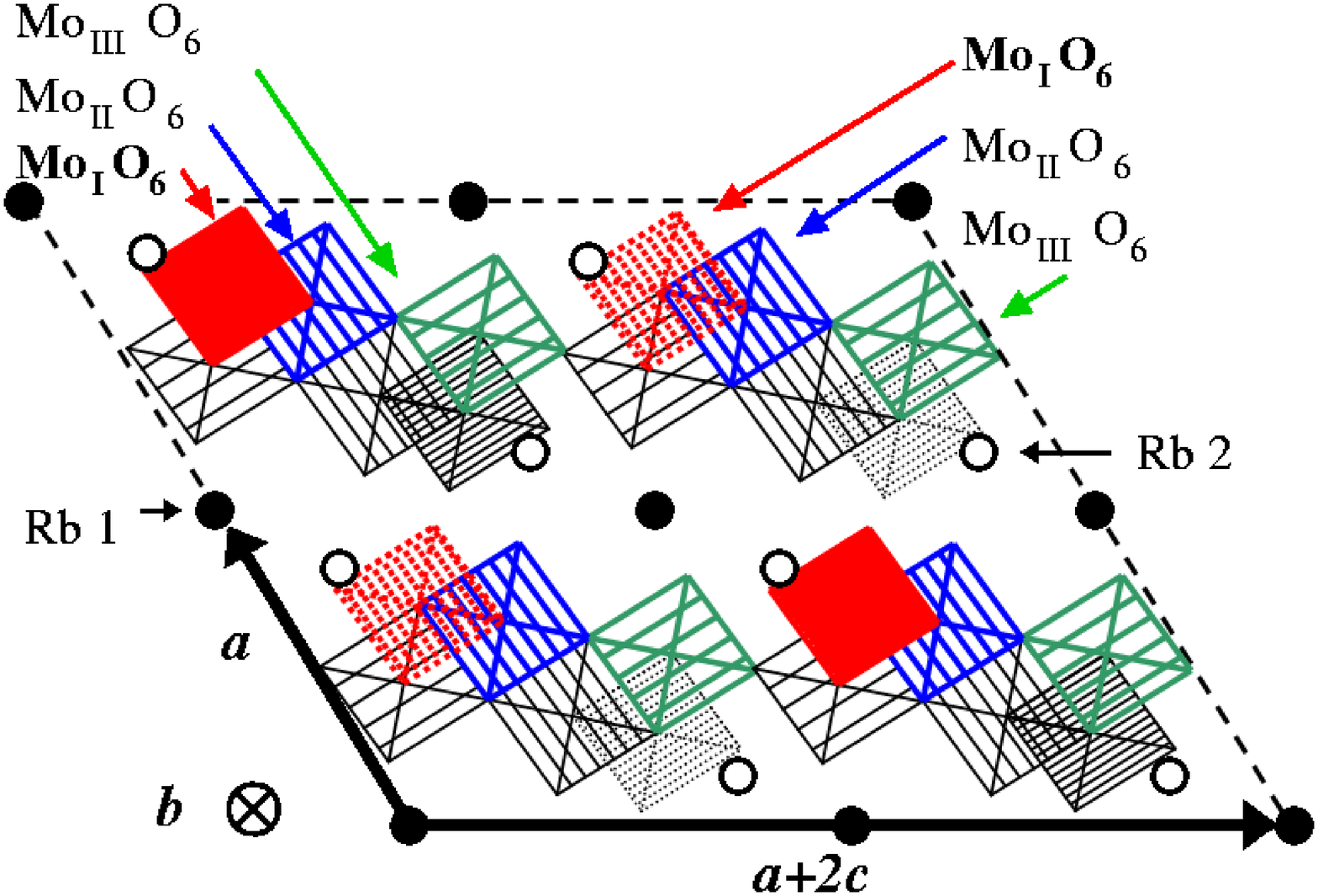}
}
\caption{(Color online) (a) Idealized surface structure of $\mbox{Rb}_{0.3}\mbox{MoO}_3$ in the (\=201) plane. (b) Idealized side view of the room temperature $\mbox{Rb}_{0.3}\mbox{MoO}_3$  structure projected onto the plane perpendicular to the $\bm b$ axis. Closed circles are Rb atoms at the uppermost positions of the ``surface'' (labeled 1) and empty circles are Rb atoms 1.2 \AA~below (labeled 2). The three highest octahedra with respect to the ``surface'' are the dashed squares indicated by the arrows. Their centers lie at levels 1.8, 2.4 and 3.5 \AA~below the ``surface''.}
\label{structure} 
\vspace*{-0.45cm}
\end{figure}

Low-dimensional molybdenum and tungsten oxides and bronzes have been the focus of much attention because of the charge density wave (CDW) and associated phenomena they exhibit \cite{reviews}. The blue bronzes, $\mbox{A}_{0.3}\mbox{MoO}_3$ (A = K, Rb, Tl), are quasi-one-dimensional metals exhibiting a metal to insulator transition and are among the most intensely studied of these materials. Their crystal structure is built from Mo$\mbox{O}_{3}$ layers in between which the cations reside (Fig. \ref{structure})\cite{graham}. Despite many attempts, observation of the CDW in these materials by scanning tunneling microscopy (STM) has been elusive. Only very recently, high resolution STM images of an in-situ cleaved (\={2}01) surface of the rubidium blue bronze, $\mbox{Rb}_{0.3}\mbox{MoO}_3$, have been obtained at low-temperature and in ultra-high vacuum (UHV)\cite{brun}. Both the molecular lattice and the CDW superlattice were observed simultaneously at temperatures well below the CDW transition temperature.\\
\hspace*{0.5cm}Comparison of these images with previous first-principles density functional theory (DFT) calculations for the bulk\cite{mozos} as well as with experimental information of the bulk structure of the modulated phase\cite{schutte} is quite puzzling. For instance, some of the most intense features of the STM images are associated with the  $\mbox{Mo}_{\rm I}\mbox{O}_{6}$ octahedra (Fig. \ref{structure}), which are only weakly involved in the CDW transition according to the superlattice structural study\cite{schutte}. Parenthetically, the $\mbox{Mo}_{\rm I}$ orbitals have a minor contribution to the states near the Fermi level according to the first-principles calculations\cite{mozos}. These and related observations prompted the present work. Here we report a first-principles study of the STM images of modulated and non modulated $\mbox{Rb}_{0.3}\mbox{MoO}_3$ with special emphasis on the influence of the alkali atoms at the surface. This is a key issue when trying to directly observe the CDW modulations in materials like low-dimensional bronzes because it may affect the surface nesting vector and hence, the nature of the modulation observed. We also report new experimental results concerning inhomogeneities in the surface modulation wave vector providing support for the analysis. Finally, a clearcut understanding of the STM images for the blue bronze emerges from this work. 
\newline
The present calculations were carried out using a numerical atomic orbitals DFT\cite{hohenberg} approach, which has been developed and designed for efficient calculations in large systems and implemented in the SIESTA code\cite{soler}. We have used the generalized gradient approximation to DFT and, in particular, the functional of Perdew, Burke and Ernzerhof\cite{perdew}. Only the valence electrons are considered in the calculation, with the core being replaced by non-local norm-conserving scalar relativistic pseudopotentials\cite{trouiller} factorized in the Kleinman-Bylander form\cite{kleinman}. Nonlinear partial-core corrections to describe the exchange and correlations in the core region were included for Mo\cite{louie}. We have used a single-$\zeta$ basis set including polarization orbitals for Mo atoms, as obtained with an energy shift of 0.02 Ry\cite{soler}. We verified that the description of bulk bands for this and related bronzes using this basis size, especially at Fermi level, is essentially the same as when using a split-valence double-$\zeta$ basis set including polarization for all atoms. The energy cutoff of the real space integration mesh was 300 Ry. Calculations for slabs of different thickness (containing from one to four octahedral layers) were carried out. For the superlattice we used the structure of reference 5 assuming a commensurate value (0.75) of the $\bm {b^*}$ component of the modulation vector. The Brillouin zone (BZ) was sampled using grids of (2x21x1) and (2x21x6) $k$-points for the slabs and the bulk, respectively\cite{monkhorst}. The energy cutoff and $k$-points values were tested against well converged values. 
\newline
\begin{figure}[ht]
\centering
\subfiguretopcapfalse
\subfigure[]{\hbox to 0.0in{\hfil\null}}%
\setcounter{subfigure}{0}
\subfiguretopcaptrue
\subfigure
{
    \label{bands-bulk}
\includegraphics*[height=4cm]{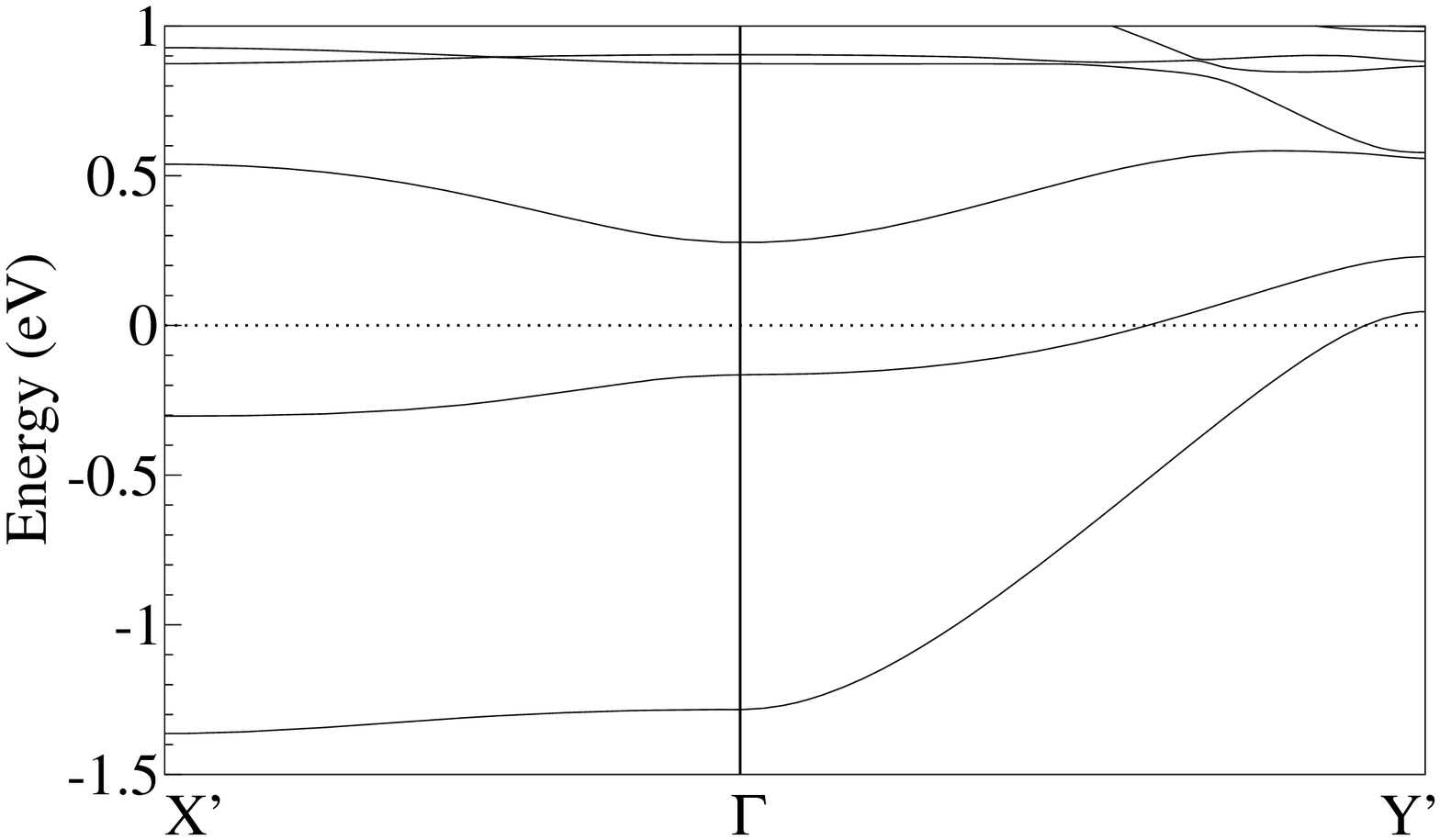}
}
\newline
\subfiguretopcapfalse
\subfigure[]{\hbox to 0.0in{\hfil\null}}%
\setcounter{subfigure}{1}
\subfiguretopcaptrue
\subfigure
{
    \label{neutra}
\includegraphics*[height=4cm]{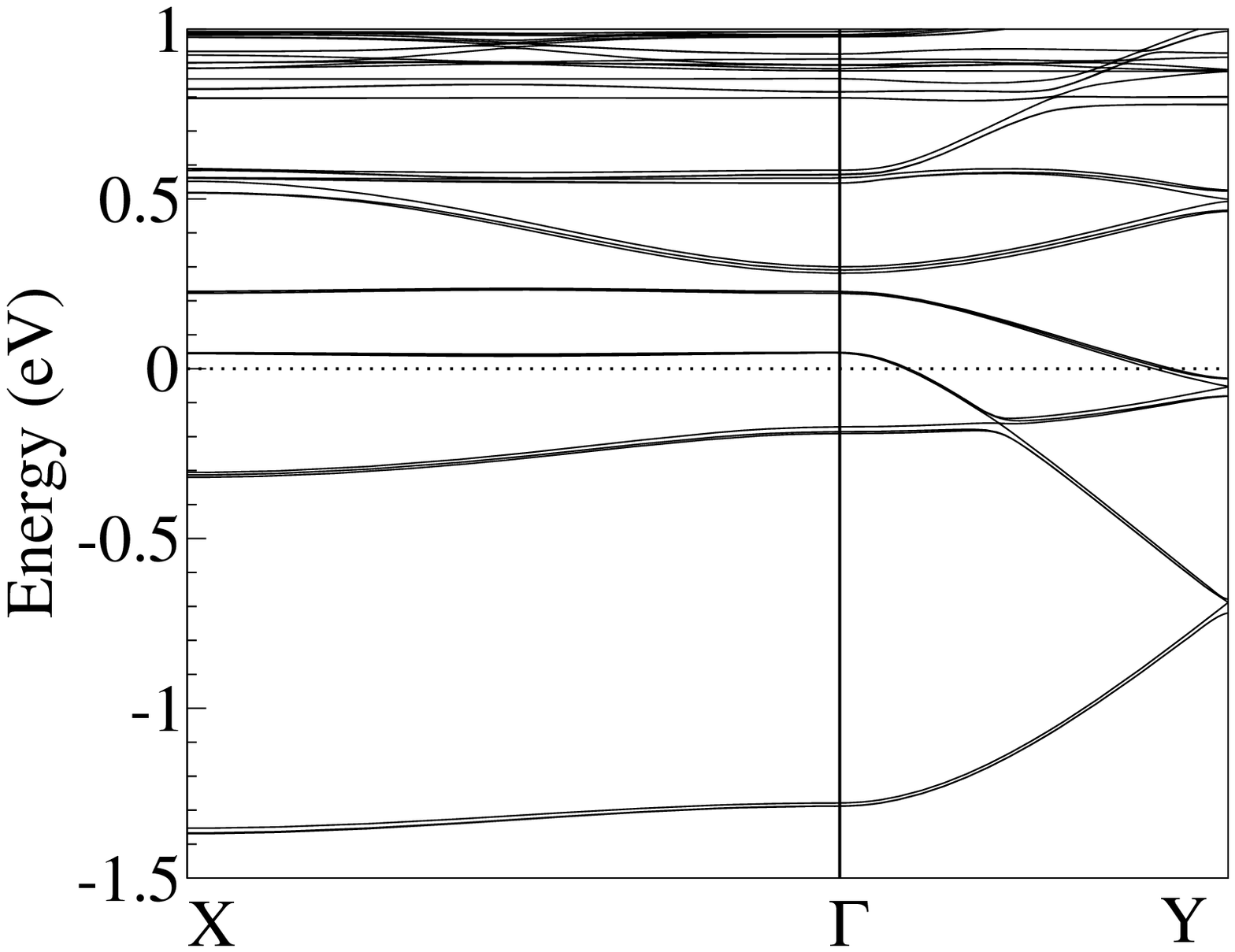}
}
\newline
\subfiguretopcapfalse
\subfigure[]{\hbox to 0.0in{\hfil\null}}%
\subfiguretopcaptrue
\setcounter{subfigure}{2}
\subfigure
{
    \label{uno-uno}
\includegraphics*[height=4cm]{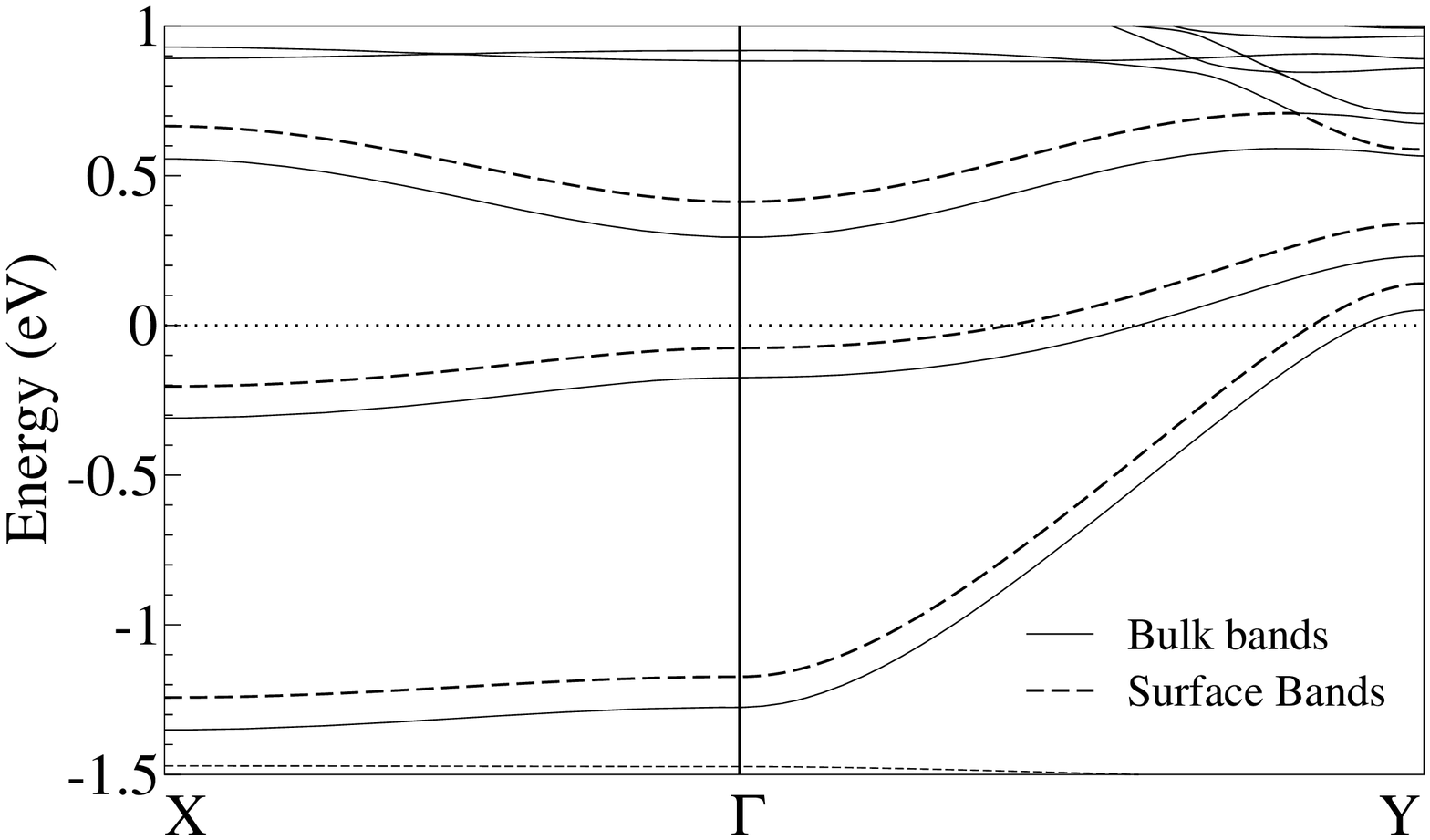}
}
\newline
\caption{Band structure for: (a) bulk $\mbox{Rb}_{0.3}\mbox{MoO}_3$; (b) a slab preserving the bulk stoichiometry at the surface, and (c) a slab with a defect of Rb atoms (one every three) at the surface. In (a)  $\Gamma = (0,0,0)$, $X'=(\frac{1}{2},0,0)$ and $Y'=(0,\frac{1}{2},0)$ in units of the $\bm {a'^{*}}$, $\bm {b'^{*}}$ and $\bm {c'^{*}}$ reciprocal lattice vectors \cite{mozos}. In (b) and (c) $\Gamma = (0,0)$, $X=(\frac{1}{2},0)$ and  $Y=(0,\frac{1}{2})$ in units of the corresponding rectangular reciprocal lattice vectors.}
\label{fig-bands} 
\vspace*{-0.45cm}
\end{figure}
The band structure for bulk $\mbox{Rb}_{0.3}\mbox{MoO}_3$ contains two partially filled bands (Fig. \ref{bands-bulk}). The CDW in this material is due to the interband nesting among these quasi-one-dimensional bands so that the CDW wave vector, $\bm {q_{CDW}}$, is given by $\bm{ k_{f1} + k_{f2}}$ where $\bm {k_{fi}}$ is the Fermi wave vector of band $\bm i$ \cite{pouget,whangbo,gweon}. Since there are three electrons per unit cell to fill these bands, a $\bm {q_{CDW}}$ component along the chain direction of $0.75\bm {b^*}$ is predicted, which is the observed value at low temperature\cite{pouget2}. In order to appropriately model the (\=201) blue bronze surface we carried out calculations for slabs including different number of octahedral (and rubidium) layers as well as different distributions and concentration of surface rubidium atoms. There are three rubidium atoms per repeat unit of a layer, two of them (type 2, empty circles in Fig. \ref{sideview}) are very near the octahedral layers and the third one (type 1, full circles in Fig. \ref{sideview}) is equidistant of the two layers. Among these interlayer Rb atoms, only the type 2 Rb atoms closest to the surface and the type 1 Rb atoms may remain at the surface after cleaving the sample. Since these atoms are expected to relax from their bulk crystallographic position, we optimized their position with respect to the surface. These positions were the basis for all remaining calculations.

\begin{figure}
\centering
\epsfxsize=5.5cm
\epsffile{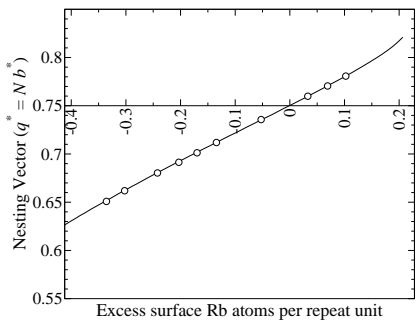}
\caption{Continuous line: calculated $\bm b^*$ component of the surface nesting vector of the (\=201) surface of rubidium blue bronze versus the density of alkali atoms at the surface. The horizontal axis indicates the number of excess Rb atoms at the surface per unit cell, with the zero corresponding to the stochiometric Rb composition. Empty circles refer to the experimental values  $\bm {q^*}$ probed by STM (See text)}
\label{nesting-vector}
\vspace*{-0.45cm}
\end{figure}

The main conclusions of our study were: i) the number of octahedral layers used in the computations is irrelevant; ii) the key factor in controlling the shape of the surface bands near the Fermi level is the number of Rb atoms at the surface. Shown in Fig. \ref{neutra} is the band structure for a surface which preserves the stoichiometry of the bulk (i.e., 1.5 Rb atoms  per repeat unit). Despite the folded-like shape of the bands, due to the fact that we used a unit cell twice larger along $\bm b$ in order to model the partial occupation of the Rb sites, the partially filled bands are nearly identical to those of the bulk. In contrast, when the repeat unit of the cell at the surface is covered by just one Rb atom, i.e., 0.5 less than in the stoichiometric case, the band structure is noticeably different (Fig. \ref{uno-uno}); the correspondig surface bands  are shifted upwards with respect to those of the bulk. Calculations for the case of an excess of Rb atoms with respect to the stoichiometric situation led to an opposite band shift. After carrying out computations for several situations we conclude that different concentrations of Rb atoms at the surface generate a nearly rigid energy shift of the surface bands with respect to those of the bulk. This result has the important implication that the nesting vector at the surface changes with the Rb content. In fact, it is possible to infer the concentration of Rb atoms at the surface that produces a given surface nesting vector, see Fig. \ref{nesting-vector}.

Let us now search for the experimental consequences of this feature. As reported in previous work\cite{brun} we have observed with careful STM experiments the nearly commensurate value of the projection of $\bm {q_{CDW}}$ onto the (\=201) plane. Hence, defining N=$\bm {q_{b^*}}/\bm{b^*}$ where $\bm {q_{b^*}}$ is the $\bm {b^*}$ component of $\bm {q_{CDW}}$, N=0.75 is the bulk reported value and 1-N=0.25 is the quantity reported by STM experiments (see part II and III in Ref. 3). However, it was mentionned that, on some optically flat terraces, 1-N was found to deviate from the 0.25 value yielding inhomogeneities for $\bm {q_{b^*}}$. We report here experimental results concerning the inhomogeneities of $\bm {q_{b*}}$ measured by STM in rubidium blue bronze. Three in-situ cleaved $\mbox{Rb}_{0.3}\mbox{MoO}_3$ samples from the same batch were investigated with several mechanically sharpened Pt/Ir tips. All samples were prepared in a very similar manner. They were cleaved at room temperature in UHV and rapidly introduced into the cold STM head. All the STM measurements consisted of constant current mode topographical images and were performed at 77K or at 63K, well below the transition temperature (T$_c$=180K). Optically large and flat plateaus were carefully selected to perform our measurements. Typical experimental conditions were $\pm$450mV for the applied bias voltage and from 50 to 150pA for the tunneling current. Much care was taken in order to achieve molecular resolution and CDW resolution with scanning areas ranging from $20\times20$nm$^2$ to $50\times50$nm$^2$. This was necessary to allow accurate measurements of the 1-N ratio for a typical $512\times512$ pixels image resolution. This 1-N ratio was directly extracted  from the 2D Fourier transform of the STM image. At a given tip location several mesurements were always performed to ensure that the measured 1-N value was reproducible within an error bar of 10\%.

It was found that optically distinct plateaux (of at least several 100 $\mu$m$^2$ area) could yield distinct values of 1-N significantly different from the 0.25 bulk value. Moreover, on the same plateau, different locations estimated to be at least several $\mu$m from each other yielded differences in 1-N values that were much greater than the typical error bar for a single location 1-N measurement, leading to clear inhomogeneities of the surface $\bm {q_{CDW}}$ wave vector. On the contrary, displacements along $\bm b$ or $\bm {a+2c}$ on the scale of tens of nanometers from a given position of measurement, did not lead to noticeable changes of $\bm {q_{b^*}}$. This shows that all measurements were performed far enough from CDW domain boundaries. On the same plateau the greatest change in 1-N value ranged from 0.21 to 0.32. As a result of our study the averaged 1-N values were found in the range 0.21 to 0.35, as indicated by the empty circles in Fig. \ref{nesting-vector} for the N value. According to the present calculation these inhomogeneities would correspond to an excess of surface alkali atoms ranging from about 0.1 to -0.35 per repeat unit. These predictions would then be consistent with the hypothesis that the distribution of type 1 alkali atoms is responsible for the experimentally reported inhomogeneities. These inhomogeneities are consistent with photoemission and grazing incidence x-ray diffraction results \cite{fedorov,zhu}, which both reported that about 100K, N almost equals 0.75. This is because STM probes $\bm {q_{CDW}}$ only at the uppermost layer of the compound and very locally at the nanometer scale inside a single CDW domain. On the other hand both x-ray and photoemission experiments probe deeper layers, which remain unaffected by the inhomogeities present in the first layer according to the present calculations, and over a much larger in-plane scale. This results in an averaging value of $\bm {q_{CDW}}$ which does not show any surface inhomogeneities.
\begin{figure}
\centering
\subfiguretopcapfalse
\subfigure[]{\label{stm-exp}\hbox to 0.1in{\hfil\null}}%
\subfiguretopcaptrue
\subfigure
{
\includegraphics*[height=4cm]{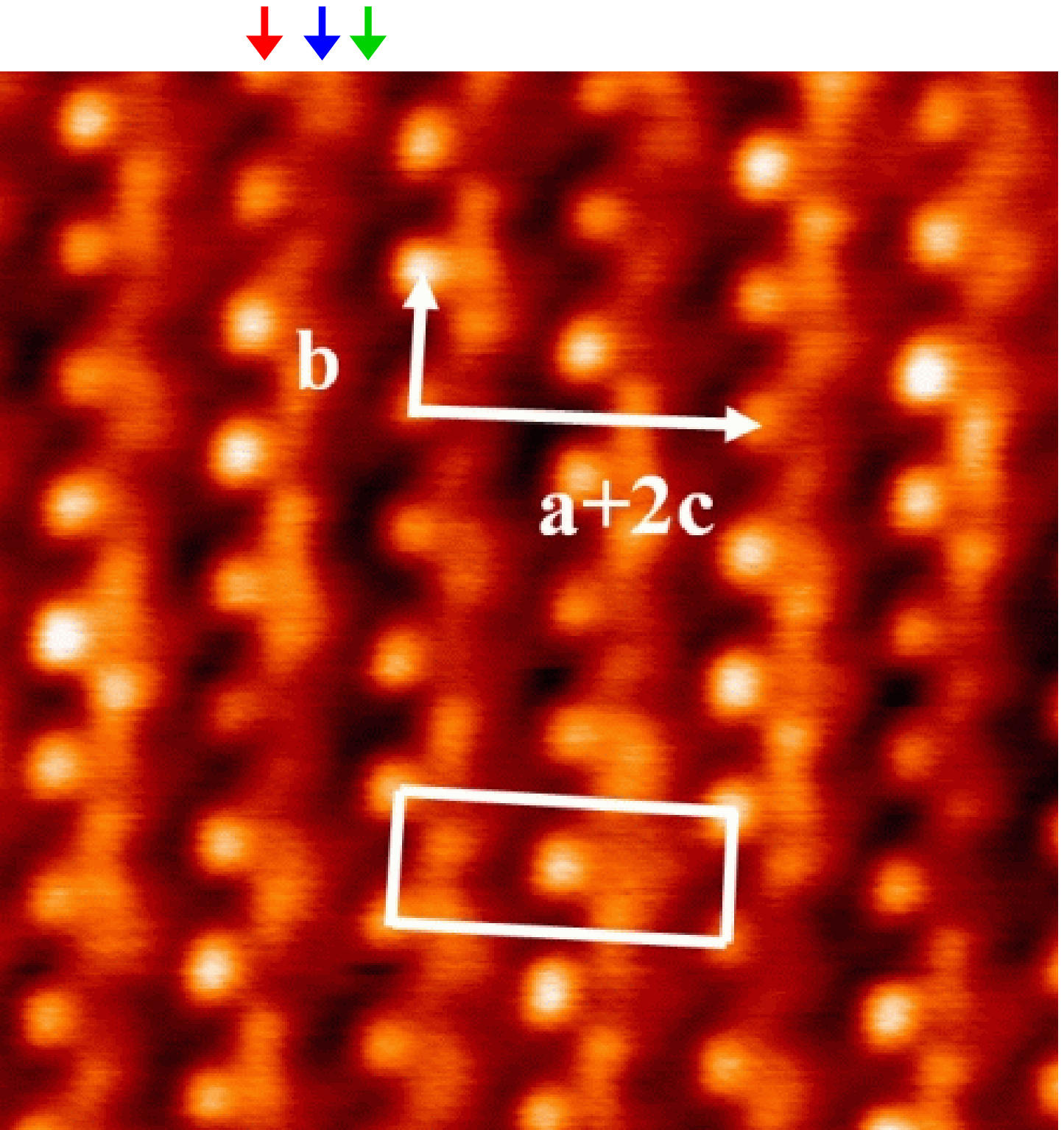}
\includegraphics*[width=4cm]{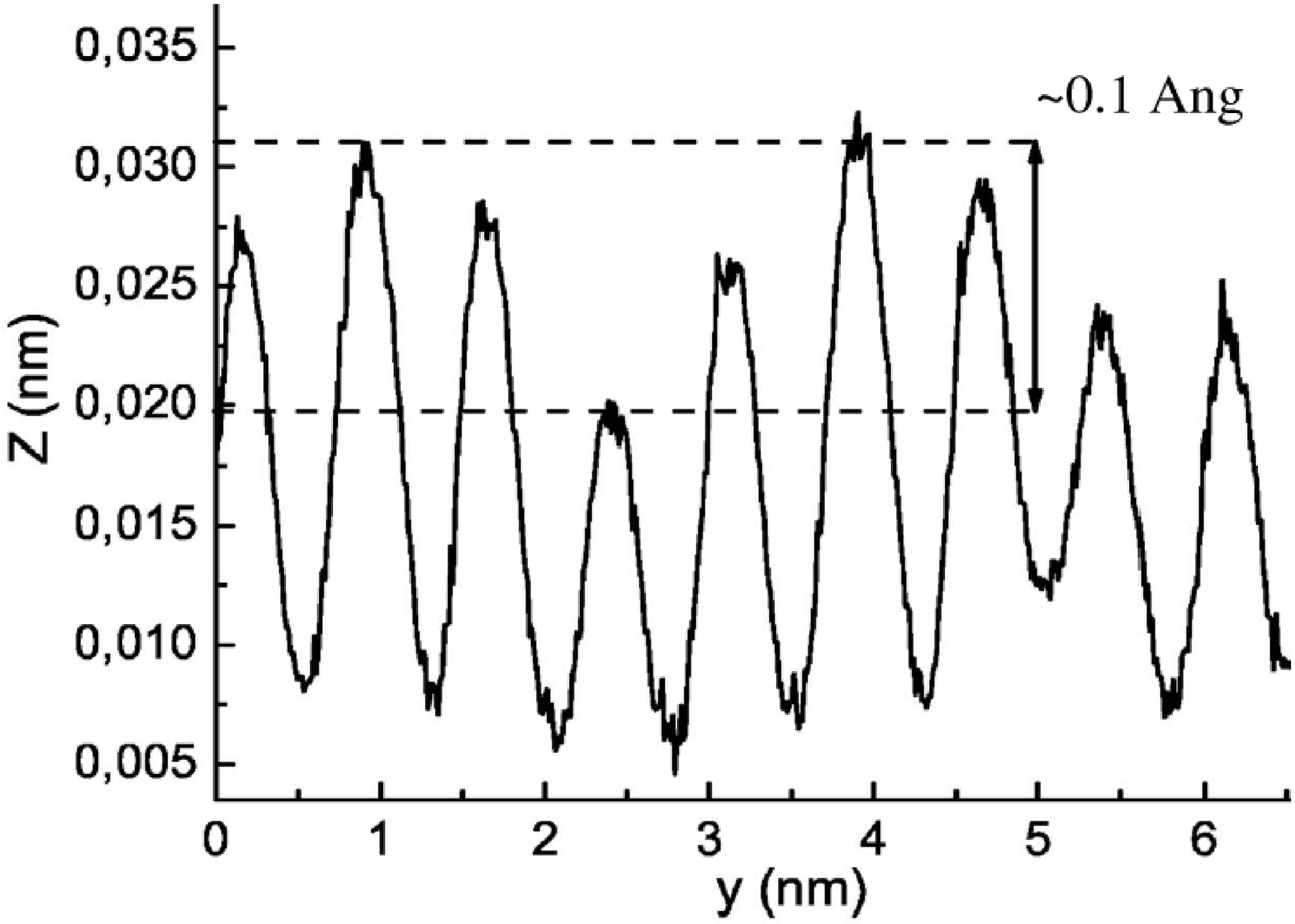}
}
\setcounter{subfigure}{1}
\subfiguretopcapfalse
\subfigure[]{\label{stm-teo}\hbox to 0.1in{\hfil\null}}%
\subfiguretopcaptrue
\subfigure
{
\includegraphics*[height=4cm]{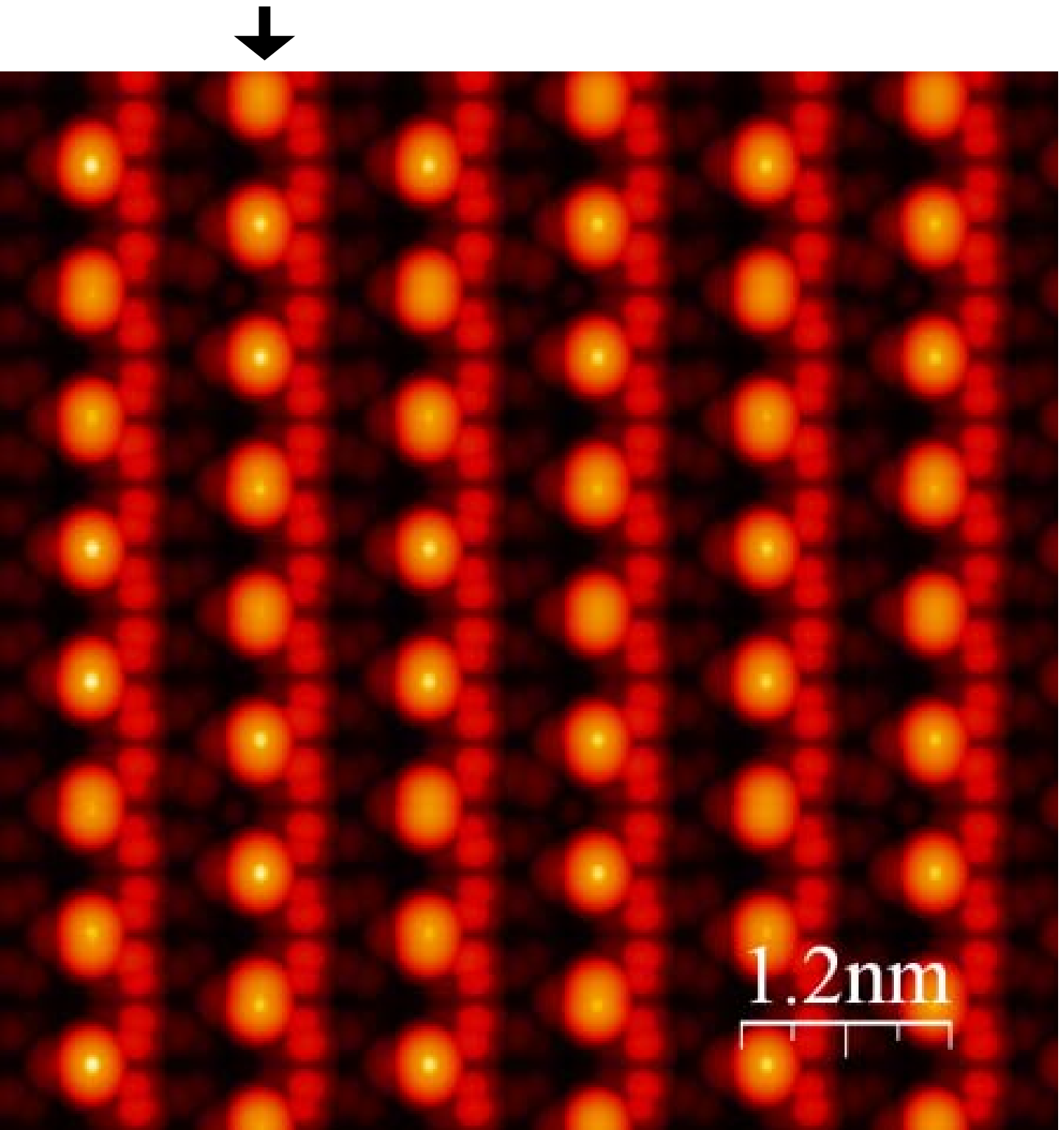}
\includegraphics*[width=4cm]{prof-teo.eps}
}
\caption{(Color online) (a) Constant current mode topographical image of $6.2\times6.2$ nm$^2$ on (\=201) plane of $\mbox{Rb}_{0.3}\mbox{MoO}_3$ at 63K (raw data image). The applied bias voltage is 420 mV and the set-up tunneling current is 110 pA. Molecular lattice and CDW superlattice coexist in the image. The three arrows indicate respectively the observed type I and II $\mbox{Mo}\mbox{O}_{6}$ octahedra and the expected position of the $\mbox{Mo}_{\rm III}\mbox{O}_{6}$ octahedra. Associated profile along $\mbox{Mo}_{\rm I}\mbox{O}_{6}$ octahedra idicated by left arrow (from Ref. \cite{brun}). (b) (Color online) Calculated image and associated profile along the $\mbox{Mo}_{\rm I}\mbox{O}_{6}$ octahedra for the modulated phase of $\mbox{Rb}_{0.3}\mbox{MoO}_3$.}
\label{stm-comparison}
\end{figure}

\begin{figure}
\centering
\epsfxsize=8.5cm
\epsffile{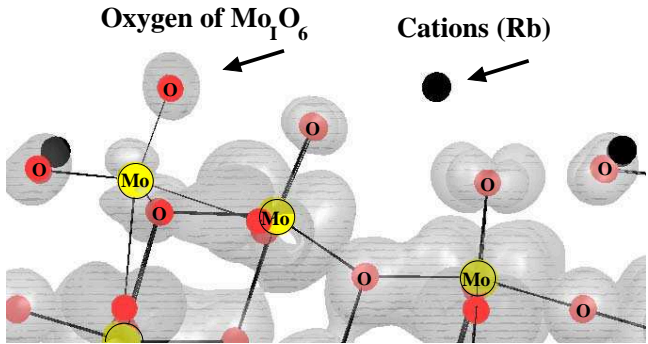}
\caption{(Color online) Iso-Charge density (side view as in Fig. \ref{sideview}) integrated from the Fermi level to 0.5 eV above. In the image in color the Rb, Mo and O atoms are shown as black, yellow and red balls, respectively.}
\label{density-ldos}
\vspace*{-0.45cm}
\end{figure}
We now turn to the analysis of the main features of the STM images for the modulated phase of $\mbox{Rb}_{0.3}\mbox{MoO}_3$ in stoichiometric conditions. As shown in Fig. \ref{stm-comparison}, where the chains along $\bm {b}$ are readily visible, there is a very good agreement between the experimental and calculated images. The observed STM pattern inside the surface elementary unit cell consists of one well defined ball next to a more elongated continuous pattern along $\bm {b}$ and was attributed\cite{brun} to the $\mbox{Mo}_{\rm I}\mbox{O}_{6}$ and $\mbox{Mo}_{\rm II}\mbox{O}_{6}$ octahedra respectively.
In order to better understand these images we report in Fig. \ref{density-ldos} a plot of the iso-charge density integrated from the Fermi level to 0.5 eV above. Two features must be noted. First, there is essentially no contribution of the Rb atoms. This provides computational support for the suggestion that the STM measurement is not sensitive to them\cite{walter}. Second, the density is noticeable around the outer O atom of the $\mbox{Mo}_{\rm I}\mbox{O}_{6}$ octahedra. These O atoms are the uppermost part of the surface, those of the $\mbox{Mo}_{\rm II}\mbox{O}_{6}$ octahedra staying approximately 0.6 \AA~below. These two facts together easily explain why the brightest spots originate from the outer O atom of the $\mbox{Mo}_{\rm I}\mbox{O}_{6}$ octahedra. In addition, the amplitude of the bulk vertical displacement of the outer O atom of the $\mbox{Mo}_{\rm I}\mbox{O}_{6}$ octahedra in the modulated structure used in the calculation is between three and four times smaller than the amplitude of the calculated density profile along the $\mbox{Mo}_{\rm I}\mbox{O}_{6}$ octahedra (Fig.  \ref{stm-teo}). This means that the STM experiment is mostly measuring the differences in local density of states (LDOS) associated with these O atoms as a result of the existence of the CDW and not the differences in height of these atoms. The above mentioned apparent contradiction with previous works is solved by the results of Fig. \ref{density-ldos}. Around 42.5\% of the charge density in this figure is associated to $\mbox{Mo}_{\rm II}$, 22.1\% to $\mbox{Mo}_{\rm III}$, 1.9\% to $\mbox{Mo}_{\rm I}$ and 1.2\% to the outer O of the $\mbox{Mo}_{\rm I}\mbox{O}_{6}$. Consequently, the CDW modulation mostly affects the $\mbox{Mo}_{\rm II}$ and $\mbox{Mo}_{\rm III}$ atoms. However, because of the non negligible participation of the outer O atoms of the $\mbox{Mo}_{\rm I}\mbox{O}_{6}$ octahedra, as a result of the strong hybridization between the Mo and O orbitals, the orbital mixing associated with the CDW modulation affects the LDOS of these atoms, leading to the differences in the profile of Fig. \ref{stm-comparison}. According to the present results the more continuous path must originate from the outer O atoms of the $\mbox{Mo}_{\rm II}\mbox{O}_{6}$ octahedra but also from the $\mbox{Mo}_{\rm II}$ atoms which strongly participate in the wave function. However the difference in height with respect to the uppermost part of the surface leads to the considerably less intense signal. Finally, the $\mbox{Mo}_{\rm III}\mbox{O}_{6}$ octahedra, which lie considerably deeper (i.e., approximatly 1.7 \AA~lower than the $\mbox{Mo}_{\rm I}\mbox{O}_{6}$ octahedra) are not visible at all in the STM image even though the CDW modulation strongly affects them. 

In summary, a combined theoretical-experimental approach has led to an in-depth understanding of the STM observation for the CDW in quasi-one-dimensional blue bronze. The puzzling relationship between how CDW reveals in STM and x-ray experiments has been understood. The decisive role played by the surface Rb atoms, leading to experimental surface CDW wave vector inhomogeneities, has been clarified.

Work at Bellaterra was supported by DGI-Spain (Project BFM2003-03372-C03), Generalitat de Catalunya (2005 SGR 683), CSIC (I3P-BPD2001-1) and by grants for computer time from the CESCA-CEPBA. Work at Marcoussis was supported by R\'{e}gion Ile de France (SESAME project n$^{\circ}$ 1377) and by Conseil G\'{e}n\'{e}ral of Essone.

\end{document}